\documentclass[10pt]{article}

\usepackage{authblk}
\usepackage{natbib}
\usepackage{mathtools}
\usepackage{amsmath}
\usepackage{amsthm}
\usepackage{amssymb}
\usepackage{amsbsy}
\usepackage{amsfonts}
\usepackage{amscd}
\usepackage{mathrsfs}
\usepackage{bm}
\usepackage{breqn}
\usepackage{color, soul}
\usepackage{enumerate}
\usepackage{empheq}
\usepackage{rotating}
\usepackage{ftnxtra}
\usepackage{fnpos}
\usepackage[titletoc,toc,title]{appendix}
\usepackage{euscript}
\usepackage{graphicx}
\usepackage{epsfig}
\usepackage{epstopdf}
\DeclareGraphicsExtensions{.pdf,.png,.jpg,.eps}
\usepackage{pstool}
\usepackage{upgreek}
\usepackage{mathrsfs}

\usepackage{psfrag}

\numberwithin{equation}{section}

\theoremstyle{plain}	
\newtheorem{thm}{Theorem}[section]

\newtheorem{prop}[thm]{Proposition}
\newtheorem*{prop*}{Proposition} 
\theoremstyle{definition}	

\newtheorem{remark}[thm]{Remark}

\usepackage{graphicx}
\usepackage{lipsum}

\usepackage{caption}
\usepackage{subcaption}

\usepackage{hyperref}
\hypersetup{colorlinks=true, linkcolor=blue}
\hypersetup{colorlinks=true,citecolor=blue}

\DeclareMathAlphabet{\mathpzc}{OT1}{pzc}{m}{it}

\usepackage{amsmath, amsthm, amssymb}

\usepackage{cleveref}

\DeclarePairedDelimiter\abs{\lvert}{\rvert}

\makeatletter
\newsavebox{\@brx}
\newcommand{\llangle}[1][]{\savebox{\@brx}{\(\m@th{#1\langle}\)}%
  \mathopen{\copy\@brx\mkern2mu\kern-0.9\wd\@brx\usebox{\@brx}}}
\newcommand{\rrangle}[1][]{\savebox{\@brx}{\(\m@th{#1\rangle}\)}%
  \mathclose{\copy\@brx\mkern2mu\kern-0.9\wd\@brx\usebox{\@brx}}}%
\let\oldabs\abs
\def\abs{\@ifstar{\oldabs}{\oldabs*}}
\makeatother

\usepackage{accents}

\newcommand{\Fe}{\accentset{e}{\mathbf{F}}}

\newcommand{\Fa}{\accentset{a}{\mathbf{F}}}

    {\end{bmatrix}}%

\usepackage{enumitem}

\usepackage[utf8]{inputenc}
\usepackage{dutchcal}
\usepackage{multirow}

\usepackage{xcolor}

\usepackage{setspace}

\begin{document}

\title{\textbf{Universal Deformations in Compressible Isotropic Cauchy Elastic Solids with Residual Stress}}

\author[1,2]{Arash Yavari\thanks{Corresponding author, e-mail: arash.yavari@ce.gatech.edu}}
\author[3]{Jos\'e Merodio}
\author[4]{Mohd H.B.M. Shariff}

\affil[1]{\small \textit{School of Civil and Environmental Engineering, Georgia Institute of Technology, Atlanta, GA 30332, USA}}
\affil[2]{\small \textit{The George W. Woodruff School of Mechanical Engineering, Georgia Institute of Technology, Atlanta, GA 30332, USA}}
\affil[3]{\small \textit{Departamento de Matem\'atica Aplicada a las TIC, 
ETS de Ingenier\'ia de Sistemas Inform\'aticos, 
Universidad Polit\'ecnica de Madrid, 
28031 Madrid, Spain}}
\affil[4]{\small \textit{Khalifa University of Science, Technology and Research, Sharjah, United Arab Emirates}}

\maketitle

\begin{abstract}
We investigate universal deformations in compressible isotropic Cauchy elastic solids with residual stress, without assuming any specific source for the residual stress. We show that universal deformations must be homogeneous, and the associated residual stresses must also be homogeneous. Since a non-trivial residual stress cannot be homogeneous, it follows that residual stress must vanish. Thus, a compressible Cauchy elastic solid with a non-trivial distribution of residual stress cannot admit universal deformations. These findings are consistent with the results of \citet{YavariGoriely2016}, who showed that in the presence of eigenstrains, universal deformations are covariantly homogeneous and in the case of simply-connected bodies the universal eigenstrains are zero-stress (impotent).
\end{abstract}

\begin{description}
\item[Keywords:] Universal deformation, Cauchy elasticity, hyperelasticity, Green elasticity, isotropic solids, residual stress.
\end{description}


\section{Introduction}

A \emph{universal deformation} is one that can be maintained in the absence of body forces for all materials within a given class, by the application of boundary tractions alone. 
While the required boundary tractions depend on the specific material, the deformation itself does not. 
The concept of universal deformations (sometimes also referred to as ``controllable" or ``general") was introduced by Ericksen in his seminal papers \citep{Ericksen1954,Ericksen1955}, building on the earlier work of Rivlin \citep{Rivlin1948,Rivlin1949a,Rivlin1949b}. 
 He proved that in homogeneous compressible isotropic solids all universal deformations are homogeneous \citep{Ericksen1955}, while for homogeneous incompressible isotropic solids he discovered four nontrivial families of universal deformations \citep{Ericksen1954}. 
Ericksen had conjectured that deformations with constant principal invariants must be homogeneous. 
This was later shown to be false by \citet{Fosdick1966}, and subsequently a fifth family of universal deformations, consisting of inhomogeneous deformations with constant principal invariants, was discovered independently by \citet{SinghPipkin1965} and \citet{KlingbeilShield1966}.
The question of whether there are additional inhomogeneous constant-principal invariant universal deformations remains open \citep{Marris1970,Kafadar1972,Marris1975,Marris1982,Fosdick1969,Fosdick1971}. 

Universal deformations have played an important role in nonlinear elasticity and related theories. 
They have been central to semi-inverse solution methods \citep{Knowles1979,Tadmor2012,Goriely2017}, to the design of experiments for determining constitutive equations \citep{Rivlin1951,Saccomandi2001}, and to the construction of exact solutions for distributed eigenstrains and defects \citep{Wesolowski1968,Gairola1979,Zubov1997,YavariGoriely2012a,YavariGoriely2012b,YavariGoriely2013a,YavariGoriely2014,Golgoon2018,YavariGoriely2013b,YavariGoriely2015,Golgoon2017,Yavari2021Eshelby}. 
They have been used as benchmark problems in computational mechanics \citep{Dragoni1996,Saccomandi2001,Chi2015,Shojaei2018}, and in deriving effective properties for nonlinear composites \citep{Hashin1985,Lopez2012,Golgoon2021}. 

In linear elasticity, their counterparts are \emph{universal displacements} \citep{Truesdell1966,Gurtin1972,Carroll1973,Yavari2020}, whose dependence on material symmetry has been systematically studied. 
Isotropic solids admit the largest class of universal displacements, while triclinic solids admit the smallest. 
This analysis has been extended to inhomogeneous solids \citep{YavariGoriely2022}, linear Cauchy elasticity \citep{,YavariSfyris2025}, and to linear anelasticity \citep{Yavari2022Anelastic-Universality}.

Extensions of Ericksen’s framework have included anisotropy \citep{YavariGoriely2021,Yavari2021,YavariGoriely2023Universal}, implicit elasticity \citep{Yavari2024ImplicitElasticity}, Cauchy elasticity \citep{Yavari2024Cauchy}, and anelasticity \citep{YavariGoriely2016,Goodbrake2020}. 
Despite the more general constitutive structure of Cauchy elasticity, the universal deformations and universal inhomogeneities in this setting coincide with those of Green elasticity (hyperelasticity). 
In anelasticity, it has been shown that universal deformations are covariantly homogeneous \citep{YavariGoriely2016}, and in incompressible anelasticity the universal eigenstrain distributions corresponding to the six known families were identified in \citep{Goodbrake2020}. 
Recent studies have also considered accreting bodies \citep{YavariPradhan2022,YavariAccretion2023,PradhanYavari2023} and liquid crystal elastomers \citep{LeeBhattacharya2023,MihaiGoriely2023}. 

The role of other internal constraints has also received attention, particularly for fiber-reinforced solids. 
Homogeneous compressible isotropic solids reinforced with inextensible fibers were studied by \citet{Beskos1972}, who examined whether the universal deformations of incompressible isotropic solids remain universal in this setting. 
A similar study for incompressible isotropic solids was carried out in \citep{Beskos1973}. 
\citet{Beatty1978,Beatty1989} considered homogeneous compressible isotropic solids reinforced with a single family of inextensible fibers and identified the fiber distributions for which homogeneous deformations are universal. 
He showed that only three such distributions exist, and in all three cases the fibers are straight lines in both the reference and deformed configurations.
More recently, \citet{Yavari2023} studied universal displacements in compressible anisotropic linear elastic solids reinforced by a uniform distribution of inextensible straight fibers and characterized the corresponding sets for each compatible symmetry class.
\citet{Yavari2025Fibers} presented the first systematic classification of universal deformations in compressible isotropic Cauchy elastic solids reinforced by a single family of inextensible fibers. 
For straight deformed fibers, he identified a new inhomogeneous non-isochoric family, denoted \emph{Family $Z_1$}. 
He also showed that if all principal invariants are constant, only homogeneous universal deformations are possible. 
When fibers have non-vanishing curvature, the universality constraints become significantly more complex, and the existence of universal deformations in this case remains open.
Other types of internal constraints have been studied, including the inexpansibility constraint \citep{Kurashige1985}, and in-plane rigidity \citep{Tommasi1996}. 

Despite this extensive literature, the effect of \emph{residual stress} on universality has not been systematically examined. Residual stresses, which are self-equilibrated in the absence of external loads, are ubiquitous in both natural and engineered solids, arising from growth, plastic deformation, thermal processes, or other anelastic mechanisms \citep{Biot1939,Hoger1985,JohnsonHoger1995,Shams2011,MerodioOgdenRodriguez2013,MerodioOgden2016,Goriely2017}. They modify the constitutive equations and raise the question of whether universal deformations can exist in their presence. In this paper, we study universal deformations in compressible isotropic Cauchy elastic solids with residual stress. We show that universal deformations are necessarily homogeneous and that universal residual stresses must also be homogeneous. 
Because nontrivial residual stresses are necessarily inhomogeneous, homogeneous residual stresses must vanish. Therefore, a compressible isotropic Cauchy elastic solid with a nontrivial residual stress distribution cannot admit universal deformations, consistent with the results of \citet{YavariGoriely2016} for eigenstrains, where universal eigenstrains were shown to be zero-stress (impotent).

This paper is organized as follows. In \S\ref{C-Elasticity}, Cauchy elasticity is briefly reviewed. The work of \citet{YavariGoriely2016} is extended to Cauchy anealsticity in \S\ref{C-Elasticity-U-S-I}. Universal deformations of compressible isotropic Cauchy elasticity with residual stress are characterized in \S\ref{C-Elasticity-UC}. Conclusions are given in \S\ref{Sec:Conclusions}.

\section{Cauchy Elasticity} \label{C-Elasticity}

Let us consider a body whose undeformed configuration is identified with an embedded submanifold $\mathcal{B}$ of the Euclidean ambient space $\mathcal{S}$. The flat metric of the ambient space is denoted by $\mathbf{g}$, and the induced metric on the reference configuration is $\mathbf{G} = \mathbf{g}\big|_{\mathcal{B}}$. 
A deformation is a smooth map $\varphi:\mathcal{B} \to \mathcal{C} \subset \mathcal{S}$, where $\mathcal{C} = \varphi(\mathcal{B})$ denotes the deformed configuration. The tangent map of $\varphi$ defines the deformation gradient $\mathbf{F} = T\varphi$, a metric-independent linear map $\mathbf{F}(X): T_X\mathcal{B} \to T_{\varphi(X)}\mathcal{C}$ at each material point $X \in \mathcal{B}$. 
In local coordinates $\{X^A\}$ on $\mathcal{B}$ and $\{x^a\}$ on $\mathcal{C}$, the deformation gradient has components $F^a{}_A = \partial \varphi^a/\partial X^A$ and the representation $\mathbf{F}=F^a{}_A\,\frac{\partial}{\partial x^a}\otimes dX^A$, where $\{\frac{\partial}{\partial x^a}\}$ and $\{dX^A\}$ are the coordinate bases for the tangent space $T_x\mathcal{C}$ and the cotangent space $T^*_X\mathcal{B}$, respectively \citep{Abraham2012Manifolds}. 
With respect to a  local coordinate chart $\{x^a\}$ on $\mathcal{C}$, $g_{ab}$ are components of the spatial metric $\mathbf{g}$. The inverse spatial metric $\mathbf{g}^{-1}=\mathbf{g}^\sharp$ has components $g^{ab}$ such that $g^{am} g_{mb}=\delta^a_b$.\footnote{A metric $\mathbf{g}$ induces the isomorphism $\sharp:T^*\mathcal{C}\to T\mathcal{C}$, which is called the sharp operator and is used for raining indices. For example, given a $1$-form (a co-vector) $\boldsymbol{\alpha}=\alpha_a\,dx^a$, $\boldsymbol{\alpha}^\sharp=g^{ab}\,\alpha_b\,\frac{\partial}{\partial x^a}$ is its corresponding vector. Similarly, $\mathbf{g}$ induces the isomorphism $\flat:T\mathcal{C}\to T^*\mathcal{C}$, which is called the flat operator and is used for lowering indices. For example, given a vector $\mathbf{u}=u^a\,\frac{\partial}{\partial x^a}$, $\mathbf{u}^\flat=g_{ab}\,u^b\,dx^a$ is its corresponding $1$-form (co-vector).} Similarly, With respect to a  local coordinate chart $\{X^A\}$ on $\mathcal{B}$, $G_{AB}$ are components of the material metric $\mathbf{G}$. The inverse material metric $\mathbf{G}^{-1}=\mathbf{G}^\sharp$ has components $G^{AB}$ such that $G^{AM} G_{MB}=\delta^A_B$.

The transpose of deformation gradient $\mathbf{F}^{\mathsf{T}}$ has components $(F^{\mathsf{T}})^A{}_a = g_{ab}\, F^b{}_B\, G^{AB}$. 
The right Cauchy-Green strain tensor is defined by $\mathbf{C} = \mathbf{F}^{\mathsf{T}}\mathbf{F}$, with components $C^A{}_B = (F^{\mathsf{T}})^A{}_a\, F^a{}_B$. This gives the standard expression
\begin{equation}
	C_{AB} = (g_{ab} \circ \varphi)\, F^a{}_A\, F^b{}_B\,,
\end{equation}
which shows that the right Cauchy-Green strain is the pull-back of the spatial metric, i.e., $\mathbf{C}^\flat = \varphi^* \mathbf{g}=\mathbf{F}^\star\mathbf{g}\mathbf{F}$, where the flat operator is defined via the reference metric $\mathbf{G}$, and $\mathbf{F}^\star$ is dual of the deformation gradient (a metric-independent two-point tensor) defined as $\mathbf{F}^\star=F^a{}_A\, dX^A\otimes \frac{\partial}{\partial x^a}$. 

The left Cauchy-Green strain tensor is defined as $\mathbf{B}^{\sharp} = \varphi^*(\mathbf{g}^{\sharp})=\mathbf{F}^{-1}\mathbf{g}^\sharp\mathbf{F}^{-\star}$, with components $B^{AB} = F^{-A}{}_a\, F^{-B}{}_b\, g^{ab}$. The spatial counterpart of $\mathbf{C}^\flat$ is $\mathbf{c}^\flat = \varphi_*\mathbf{G}=\mathbf{F}^{-\star}\mathbf{G}\mathbf{F}^{-1}$, whose components are $c_{ab} = F^{-A}{}_a\, F^{-B}{}_b\, G_{AB}$. Likewise, the spatial analogue of $\mathbf{B}^{\sharp}$ is $\mathbf{b}^{\sharp} = \varphi_*(\mathbf{G}^{\sharp})=\mathbf{F}\mathbf{G}^\sharp\mathbf{F}^{\star}$, with components $b^{ab} = F^a{}_A\, F^b{}_B\, G^{AB}$. Note that $\mathbf{b} = \mathbf{c}^{-1}$. For more details on the geometric formulation of elasticity see \citep{MarsdenHughes1994,Yavari2019Cloaking}.

The tensors $\mathbf{C}$ and $\mathbf{b}$ share the same principal invariants $I_1$, $I_2$, and $I_3$, which are defined as follows \citep{Ogden1984,MarsdenHughes1994}:
\begin{equation} \label{Principal-Invariants}
	I_1 = \operatorname{tr}\mathbf{b} = b^{ab} g_{ab}\,, \qquad
	I_2 = \frac{1}{2}\left(I_1^2 - \operatorname{tr} \mathbf{b}^2\right)
	= \frac{1}{2}\left(I_1^2 - b^{ab} b^{cd} g_{ac} g_{bd}\right)\,, \qquad
	I_3 = \det \mathbf{b}\,.
\end{equation}

In Cauchy elasticity, the stress at a material point and at a given instant depends explicitly on the strain at that same point and time \citep{Cauchy1828,Truesdell1952,TruesdellNoll2004,YavariGoriely2025}. However, the existence of an energy function is not guaranteed.\footnote{It is important to emphasize that Cauchy elasticity does not describe all elastic materials. A more general class of elastic solids are described by implicit constitutive relations of the form $\boldsymbol{\mathsf{f}}(\boldsymbol{\sigma},\mathbf{b})=\mathbf{0}$ \citep{Morgan1966,Rajagopal2003,Rajagopal2007}. Cauchy elasticity is a special case within this broader class.} In terms of the first Piola-Kirchhoff stress tensor \citep{Truesdell1952,TruesdellNoll2004,Ogden1984}, one has
\begin{equation} \label{P-Cauchy}
	\mathbf{P}=\hat{\mathbf{P}}(X,\mathbf{F},\mathbf{G},\mathbf{g})\,.
\end{equation}
Objectivity implies that the second Piola-Kirchhoff stress must take the form \citep{TruesdellNoll2004}
\begin{equation} \label{S-Cauchy}
	\mathbf{S}=\hat{\mathbf{S}}(X,\mathbf{C}^\flat,\mathbf{G})\,.
\end{equation}
For isotropic materials, one arrives at the classical representation \citep{RivlinEricksen1955,Wang1969,Boehler1977}
\begin{equation}
	\mathbf{S}=\Lambda_0 \mathbf{G}^\sharp+\Lambda_1 \mathbf{C}^\sharp
	+\Lambda_{-1} \mathbf{C}^{-\sharp}\,,
\end{equation}
where $\Lambda_i=\Lambda_i(X,I_1,I_2,I_3)$, $i=-1,0,1$, and $\sharp$ denotes the sharp operator associated with the metric $\mathbf{G}$. 

The Cauchy stress for compressible isotropic Cauchy elastic materials admits the representation
\begin{equation}
	\boldsymbol{\sigma} =\alpha\, \mathbf{g}^\sharp+\beta \mathbf{b}^\sharp
	+\gamma\mathbf{c}^\sharp\,, 
\end{equation}
where $\alpha=\alpha(I_1,I_2,I_3)$, $\beta=\beta(I_1,I_2,I_3)$, and $\gamma=\gamma(I_1,I_2,I_3)$ are arbitrary constitutive functions, where the principal invariants are given in \eqref{Principal-Invariants}.

\subsection{Cauchy elastic solids with distributed eigenstrains} 

An elastic body with eigenstrains is residually-stressed, in general. Deformation gradient is multiplicatively decomposed as $\mathbf{F}=\Fe\Fa$, where $\Fe$ and $\Fa$ are the elastic and anelastic distortions, respectively.\footnote{See \citep{SadikYavari2017, yavariSozio2023} for more details on this decomposition and its history.} These distortions are not compatible, in general.
Imagine that the body in its reference configuration is partitioned into a large number of small (infinitesimal) pieces. The anelastic distortion $\Fa$ maps each small piece to its local relaxed state.
Let us consider an infinitesimal line element in the reference configuration. It is represented as $\mathbf{N} dS$, where $\mathbf{N}$ is a unit vector. $\Fa\mathbf{N} dS$ is relaxed line element.
Suppose the Euclidean metric on the body is $\mathring{\mathbf{G}}$. The squared length of the relaxed line element is calculated as
\begin{equation}
	\llangle \Fa\mathbf{N} dS,\Fa\mathbf{N} dS\rrangle_{\mathring{\mathbf{G}}}
	=\llangle \mathbf{N} dS, \mathbf{N} dS\rrangle_{\Fa^*\mathring{\mathbf{G}}}\,,
\end{equation}
where $\mathbf{G}=\Fa^*\mathring{\mathbf{G}}=\Fa^\star\mathring{\mathbf{G}}\Fa$ is called the material metric. It is non-flat, i.e., it has non-vanishing Riemann curvature, in general. This idea goes back to \citet{Eckart1948} and \citet{Kondo1949}.

In the presence of eigenstrains, the constitutive equations of a Cauchy anelastic solid has the form \eqref{P-Cauchy} or \eqref{S-Cauchy} when $\mathbf{G}$ is the material metric, see \citep{YavariGoriely2025} for more details.

\subsection{Cauchy elastic solids with residual stress} 

Let us consider a body composed of a compressible Cauchy elastic solid that, in the absence of residual stress, is isotropic.  
The second Piola-Kirchhoff and Cauchy stress tensors are denoted by $\mathbf{S}$ and $\boldsymbol{\sigma}$, respectively.  
We assume that the body is residually stressed and denote the second Piola-Kirchhoff residual stress by $\mathring{\mathbf{S}}$.  
Residual stress is self-equilibrated in the absence of body forces and boundary tractions; thus, $\operatorname{Div} \mathring{\mathbf{S}} = \mathbf{0}$ in $\mathcal{B}$ and $\mathring{\mathbf{S}}\mathbf{N} = \mathbf{0}$ on $\partial\mathcal{B}$,  
where $\mathbf{N}$ is the unit normal to the undeformed boundary, and $\operatorname{Div}$ is the divergence operator with respect to the material metric $\mathbf{G}$.  
It should be emphasized that the divergence with respect to $\mathbf{G}$ arises solely in the context of a residual stress field.  
When writing the equilibrium equations in terms of the second Piola-Kirchhoff stress, the divergence operator is taken with respect to the metric $\mathbf{C}^\flat = \varphi^* \mathbf{g}$.  
We denote the push-forward of $\mathring{\mathbf{S}}$ to the current configuration by $\mathring{\boldsymbol{\sigma}}$, i.e., $\mathring{\boldsymbol{\sigma}} = \varphi_* \mathring{\mathbf{S}} = \mathbf{F} \mathring{\mathbf{S}} \mathbf{F}^\star$.  
Although the expression for $\mathring{\boldsymbol{\sigma}}$ resembles that of the Kirchhoff stress, it is not the actual Kirchhoff stress arising from a constitutive law.  
It is simply the spatial representation of a pre-existing residual stress field, independent of the deformation.

\begin{remark}
Consider the inclusion map $\iota:\mathcal{B}\hookrightarrow\mathcal{S}$. 
The residual stress $\mathring{\mathbf{S}}$ induces a corresponding \textit{Cauchy residual stress} 
defined by $\boldsymbol{\sigma}_0=\iota_*\mathring{\mathbf{S}}$. 
With a slight abuse of notation, we identify these two tensors and simply work with $\mathring{\mathbf{S}}$.
\end{remark}

The second Piola-Kirchhoff stress $\mathbf{S}$ is an isotropic function of $\mathbf{C}$ and $\mathring{\mathbf{S}}$.  
For a symmetric second-order tensor $\mathbf{S}$ that depends isotropically on two symmetric tensors, $\boldsymbol{\mathsf{A}}=\mathbf{C}^\sharp$ and $\boldsymbol{\mathsf{B}}=\mathring{\mathbf{S}}$, its representation can be expressed as \citep{Spencer1970,Boehler1987}
\begin{equation}
	\mathbf{S}=\sum_{j=1}^m \gamma_j\,\mathbf{G}_j\,,\qquad  \gamma_j=\gamma_j(I_1,\cdots,I_n)\,,
\end{equation}
where $\{I_1,\cdots,I_n\}$ is a set of polynomial scalar invariants that is assumed to be irreducible, i.e., no $I_k$ in the set can be written as a polynominal function of the remaining invariants, and $\{\mathbf{G}_1,\cdots,\mathbf{G}_m\}$ is a set of generating tensors. For $\mathbf{S}=\mathbf{S}(\boldsymbol{\mathsf{A}},\boldsymbol{\mathsf{B}})$, an irreducible integrity basis is
\begin{equation}
\begin{aligned}
	& I_1=\operatorname{tr}\boldsymbol{\mathsf{A}}\,,&&
	I_2=\operatorname{tr}\boldsymbol{\mathsf{A}}^2\,,&&
	I_3=\operatorname{tr}\boldsymbol{\mathsf{A}}^3\,,&&
	I_4=\operatorname{tr}\boldsymbol{\mathsf{B}}\,,&&
	I_5=\operatorname{tr}\boldsymbol{\mathsf{B}}^2\,,&&
	I_6=\operatorname{tr}\boldsymbol{\mathsf{B}}^3\,,\\
	& I_7=\operatorname{tr}\left(\boldsymbol{\mathsf{A}}\boldsymbol{\mathsf{B}}\right)\,,&&
	I_8=\operatorname{tr}\left(\boldsymbol{\mathsf{A}}^2\boldsymbol{\mathsf{B}}\right)\,,&&
	I_9=\operatorname{tr}\left(\boldsymbol{\mathsf{A}}\boldsymbol{\mathsf{B}}^2\right)\,,&&
	I_{10}=\operatorname{tr}\left(\boldsymbol{\mathsf{A}}^2\boldsymbol{\mathsf{B}}^2\right)
	\,,
\end{aligned}
\end{equation}
and a set of generating tensors is
\begin{equation}
\begin{aligned}
	\left\{
	\mathbf{G}^\sharp, \boldsymbol{\mathsf{A}},\boldsymbol{\mathsf{A}}^2,\boldsymbol{\mathsf{B}},
	\boldsymbol{\mathsf{B}}^2,\boldsymbol{\mathsf{A}}\boldsymbol{\mathsf{B}}
	+\boldsymbol{\mathsf{B}}\boldsymbol{\mathsf{A}},
	\boldsymbol{\mathsf{A}}^2\boldsymbol{\mathsf{B}}+\boldsymbol{\mathsf{B}}\boldsymbol{\mathsf{A}}^2, 
	\boldsymbol{\mathsf{A}}\boldsymbol{\mathsf{B}}^2
	+\boldsymbol{\mathsf{B}}^2\boldsymbol{\mathsf{A}}
	\right\}\,.
\end{aligned}
\end{equation}
Therefore, we have the following representation for $\mathbf{S}$:
\begin{equation}
\begin{aligned}
	\mathbf{S} &= 
	\gamma_0\mathbf{G}^\sharp
	+\gamma_1\boldsymbol{\mathsf{A}}
	+\gamma_2\boldsymbol{\mathsf{A}}^2
	+\gamma_3\boldsymbol{\mathsf{B}}
	+\gamma_4\boldsymbol{\mathsf{B}}^2
	+\gamma_5 \left(\boldsymbol{\mathsf{A}}\boldsymbol{\mathsf{B}}+\boldsymbol{\mathsf{B}}\boldsymbol{\mathsf{A}}\right)
	+\gamma_6 \left(\boldsymbol{\mathsf{A}}^2\boldsymbol{\mathsf{B}}
	+\boldsymbol{\mathsf{B}}\boldsymbol{\mathsf{A}}^2\right)
	+\gamma_7 \left(\boldsymbol{\mathsf{A}}\boldsymbol{\mathsf{B}}^2
	+\boldsymbol{\mathsf{B}}^2\boldsymbol{\mathsf{A}}\right)
	\,.
\end{aligned}
\end{equation}
From the Cayley-Hamilton theorem, the square of a symmetric second-order tensor can be expressed as a linear combination of the tensor itself, its inverse, and the metric tensor $\mathbf{G}^\sharp$. This follows because, for a second-order tensor $\mathbf{A}$, the Cayley-Hamilton theorem implies that $\mathbf{A}^2$ lies in the span of $\{\mathbf{G}^\sharp, \mathbf{A}, \mathbf{A}^{-1}\}$, with coefficients depending only on the principal invariants of $\mathbf{A}$\,.
Therefore, the following is a set of generating tensors for $\mathbf{S}$
\begin{equation}
\begin{aligned}
	\left\{
	\mathbf{G}^\sharp, \mathbf{C}^\sharp,\mathbf{B}^\sharp,\mathring{\mathbf{S}},\mathring{\mathbf{S}}^2,
	\mathbf{C}\mathring{\mathbf{S}}+\mathring{\mathbf{S}}\mathbf{C},
	\mathbf{B}\mathring{\mathbf{S}}+\mathring{\mathbf{S}}\mathbf{B}, 
	\mathbf{C}\mathring{\mathbf{S}}^2+\mathring{\mathbf{S}}^2\mathbf{C}
	\right\}\,.
\end{aligned}
\end{equation}
Therefore,
\begin{equation} \label{S-Representation}
\begin{aligned}
	\mathbf{S} &= \beta_0\, \mathbf{G}^\sharp
	+\beta_1 \mathbf{C}^\sharp+\beta_2\,\mathbf{B}^\sharp+\beta_3\,\mathring{\mathbf{S}}+\beta_4\, \mathring{\mathbf{S}}^2
	+\beta_5 \big(\mathbf{C}\,\mathring{\mathbf{S}}+\mathring{\mathbf{S}}\,\mathbf{C}\big)
	+\beta_6 \big(\mathbf{B}\mathring{\mathbf{S}}+\mathring{\mathbf{S}}\mathbf{B}\big)
	+\beta_7 \big(\mathbf{C}\,\mathring{\mathbf{S}}^2+\mathring{\mathbf{S}}^2\,\mathbf{C}\big)
	\,,
\end{aligned}
\end{equation}
where the scalar-valued response functions $\beta_i = \beta_i(I_1, \cdots, I_{10})$ depend smoothly on the ten functionally independent invariants defined as \citep{Spencer1971}
\begin{equation}
\begin{aligned}
	I_1 &= \operatorname{tr} \mathbf{C} = C_{AB}\, G^{AB}\,, \\
	I_2 &=  \frac{1}{2}\left(I_1^2 - \operatorname{tr} \mathbf{C}^2\right)
	= \frac{1}{2} \left( I_1^2 - C_{AB} C_{CD}\, G^{AC}\, G^{BD} \right)\,, \\
	I_3 &= \det \mathbf{C}\,, \\
	I_4 &= \operatorname{tr} \mathring{\mathbf{S}} = \mathring{S}^{AB} G_{AB}\,, \\
	I_5 &= \operatorname{tr}(\mathring{\mathbf{S}}^2) 
	= \mathring{S}^{AC}\, \mathring{S}_C{}^B\, G_{AB}\,, \\
	I_6 &= \operatorname{tr}(\mathring{\mathbf{S}}^3) 
	= \mathring{S}^{AD}\, \mathring{S}_D{}^C\, \mathring{S}_C{}^B\, G_{AB}\,, \\
	I_7 &= \operatorname{tr}(\mathbf{C} \, \mathring{\mathbf{S}}) 
	= C^A{}_C\, \mathring{S}^{CB}\, G_{AB}\,, \\
	I_8 &= \operatorname{tr}(\mathbf{C}^2 \,\mathring{\mathbf{S}}) 
	= C^{AD}\, B_{DC}\, \mathring{S}^{CB}\, G_{AB}\,, \\
	I_9 &= \operatorname{tr}(\mathbf{C} \,\mathring{\mathbf{S}}^2) 
	= C^A{}_C\, \mathring{S}^{CD}\, \mathring{S}_D{}^B\, G_{AB}\,,\\
	I_{10} & = \operatorname{tr}(\mathbf{C}^2 \,\mathring{\mathbf{S}}^2) 
	= C^A{}_C\,C^C{}_D\,\mathring{S}^D{}_E\,\mathring{S}^E{}_B\,G^{BA}\,
	\end{aligned}
\end{equation}

The Cauchy stress $\boldsymbol{\sigma}$ is an isotropic function of $\mathbf{b}$ and $\mathring{\boldsymbol{\sigma}}$. It has the following representation 
\begin{equation} \label{Cauchy-Stress-Representation}
\begin{aligned}
	\boldsymbol{\sigma}  &= \alpha_0\, \mathbf{g}^\sharp
	+\alpha_1 \mathbf{b}^\sharp
	+\alpha_2\mathbf{c}^\sharp
	+\alpha_3\mathring{\boldsymbol{\sigma}}
	+\alpha_4 \mathring{\boldsymbol{\sigma}}^2
	+\alpha_5 \big(\mathbf{b}\mathring{\boldsymbol{\sigma}}+\mathring{\boldsymbol{\sigma}}\mathbf{b}\big)
	+\alpha_6 \big(\mathbf{c}\,\mathring{\boldsymbol{\sigma}}+\mathring{\boldsymbol{\sigma}}\,\mathbf{c}\big)
	+\alpha_7 \big(\mathbf{b}\,\mathring{\boldsymbol{\sigma}}^2+\mathring{\boldsymbol{\sigma}}^2\,\mathbf{b}\big)
	\,,
\end{aligned}
\end{equation}
where the scalar-valued response functions $\alpha_i = \alpha_i(I_1, \cdots, I_{10})$ depend smoothly on the ten functionally independent invariants, which are equivalently defined as
\begin{equation}
\begin{aligned}
	I_1 &= \operatorname{tr} \mathbf{b}^\sharp = b^{ab}\, g_{ab}\,, \\
	I_2 &= \frac{1}{2} \left( I_1^2 - b^{ab} b^{cd}\, g_{ac}\, g_{bd} \right)\,, \\
	I_3 &= \det \mathbf{b}\,, \\
	I_4 &= \operatorname{tr} \mathring{\boldsymbol{\sigma}} = \mathring{\sigma}^{ab} g_{ab}\,, \\
	I_5 &= \operatorname{tr}(\mathring{\boldsymbol{\sigma}}^2) 
	= \mathring{\sigma}^{ac}\, \mathring{\sigma}_c{}^b\, g_{ab}\,, \\
	I_6 &= \operatorname{tr}(\mathring{\boldsymbol{\sigma}}^3) 
	= \mathring{\sigma}^{ad}\, \mathring{\sigma}_d{}^c\, \mathring{\sigma}_c{}^b\, g_{ab}\,, \\
	I_7 &= \operatorname{tr}(\mathbf{b} \cdot \mathring{\boldsymbol{\sigma}}) 
	= b^a{}_c\, \mathring{\sigma}^{cb}\, g_{ab}\,, \\
	I_8 &= \operatorname{tr}(\mathbf{b}^2 \cdot \mathring{\boldsymbol{\sigma}}) 
	= b^{ad}\, b_{dc}\, \mathring{\sigma}^{cb}\, g_{ab}\,, \\
	I_9 &= \operatorname{tr}(\mathbf{b} \cdot \mathring{\boldsymbol{\sigma}}^2) 
	= b^a{}_c\, \mathring{\sigma}^{cd}\, \mathring{\sigma}_d{}^b\, g_{ab}\,, \\
	I_{10} &= \operatorname{tr}(\mathbf{b}^2 \cdot \mathring{\boldsymbol{\sigma}}^2) 
	= b^a{}_c\,b^c{}_d\,\mathring{\sigma}^d{}_e\,\mathring{\sigma}^e{}_b\,g^{ba} \,.
\end{aligned}
\end{equation}
In coordinates, the Cauchy stress is written as
\begin{equation}
\begin{aligned}
	\sigma^{ab} &= \alpha_0\, g^{ab}
	+ \alpha_1\, b^{ab}
	+ \alpha_2\, c^{ab}
	+ \alpha_3\, \mathring{\sigma}^{ab}
	+ \alpha_4\, \mathring{\sigma}^{ac}\, \mathring{\sigma}_c^{b} \\
	&\quad + \alpha_5\, \left(b^a_c\, \mathring{\sigma}^{cb} + \mathring{\sigma}^{ac}\, b_c^{b} \right)
	+ \alpha_6\, \left(c^a_c\, \mathring{\sigma}^{cb} + \mathring{\sigma}^{ac}\, c_c^{b} \right) 
	+ \alpha_7\, \left(b^a_c\, \mathring{\sigma}^{cd}\, \mathring{\sigma}_d^{b} 
	+ \mathring{\sigma}^{ac}\, \mathring{\sigma}_c^{d}\, b_d^{b} \right)
	\,.
\end{aligned}
\end{equation}

\section{Universal Deformations and Eigenstrain in Compressible Isotropic Cauchy Elasticity} \label{C-Elasticity-U-S-I}

In this section we briefly discuss universal deformations and eigenstrains in Cauchy anelasticity and extend the work of \citet{YavariGoriely2016} to Cauchy anelasticity.

When body forces are absent, the equilibrium equations take the form $\sigma^{ab}{}_{|b}=0$.\footnote{$(.)_{|a}$ denotes the covariant derivative with respect to $\frac{\partial}{\partial x^a}$. In Cartesian coordinates this reduces to a partial derivative. For any scalar field $f$, one has $f_{|a}=f_{,a}$.}  
Using the fact that the Riemannian metric is compatible with its Levi-Civita connection, i.e., $g^{ab}{}_{|c}=0$, and therefore $g^{ab}{}_{|b}=0$, we obtain  
\begin{equation} \label{Equilibrium-Compressible}
	\sigma^{ab}{}_{|b} \;=\;
	\beta\,b^{ab}{}_{|b} \,+\,\gamma\,c^{ab}{}_{|b}
	\,+\,\alpha_{,b}\,g^{ab}\,+\,\beta_{,b}\,b^{ab}\,+\,\gamma_{,b}\,c^{ab}
	\;=\;0\,.
\end{equation}
Note that
\begin{equation}
\begin{aligned}
	\alpha_{,b} &=
	\frac{\partial \alpha}{\partial I_1}\,I_{1,b}
	+ \frac{\partial \alpha}{\partial I_2}\,I_{2,b}
	+ \frac{\partial \alpha}{\partial I_3}\,I_{3,b}\,,\\
	\beta_{,b} &=
	\frac{\partial \beta}{\partial I_1}\,I_{1,b}
	+ \frac{\partial \beta}{\partial I_2}\,I_{2,b}
	+ \frac{\partial \beta}{\partial I_3}\,I_{3,b}\,,\\
	\gamma_{,b} &=
	\frac{\partial \gamma}{\partial I_1}\,I_{1,b}
	+ \frac{\partial \gamma}{\partial I_2}\,I_{2,b}
	+ \frac{\partial \gamma}{\partial I_3}\,I_{3,b}\,.
\end{aligned}
\end{equation}
Equivalently,
\begin{equation} \label{Coefficients-Derivatives}
\begin{aligned}
	\alpha_{,b} &= \alpha_1\,I_{1,b} + \alpha_2\,I_{2,b} + \alpha_3\,I_{3,b}\,,\\
	\beta_{,b} &= \beta_1\,I_{1,b} + \beta_2\,I_{2,b} + \beta_3\,I_{3,b}\,,\\
	\gamma_{,b} &= \gamma_1\,I_{1,b} + \gamma_2\,I_{2,b} + \gamma_3\,I_{3,b}\,,
\end{aligned}
\end{equation}
where
\begin{equation}
	\alpha_i = \frac{\partial \alpha}{\partial I_i}\,, 
	\qquad \beta_i = \frac{\partial \beta}{\partial I_i}\,, 
	\qquad \gamma_i = \frac{\partial \gamma}{\partial I_i}\,, 
	\qquad i=1,2,3\,.
\end{equation}
Substituting \eqref{Coefficients-Derivatives} into \eqref{Equilibrium-Compressible}, one finds
\begin{equation} 
\begin{aligned}
	& \beta\, b^{ab}{}_{|b} + \gamma\, c^{ab}{}_{|b} \\
	& + \big( I_{1,b}\, g^{ab}\, \alpha_1 + I_{2,b}\, g^{ab}\, \alpha_2 + I_{3,b}\, g^{ab}\, \alpha_3 \big) \\
	& + \big( I_{1,b}\, b^{ab}\, \beta_1 + I_{2,b}\, b^{ab}\, \beta_2 + I_{3,b}\, b^{ab}\, \beta_3 \big) \\
	& + \big( I_{1,b}\, c^{ab}\, \gamma_1 + I_{2,b}\, c^{ab}\, \gamma_2 + I_{3,b}\, c^{ab}\, \gamma_3 \big) 
	= 0 \,. 
\end{aligned}
\end{equation}
Since $\alpha$, $\beta$, and $\gamma$ are arbitrary functions, their derivatives are independent. 
Therefore, for equilibrium to hold for any compressible isotropic Cauchy anelastic solid, each coefficient must vanish separately:
\begin{empheq}[left={\empheqlbrace }]{align}
	\label{Universality-Comp-1} 
	& b^{ab}{}_{|b} = c^{ab}{}_{|b} = 0 \,,\\
	\label{Universality-Comp-2} 
	& g^{ab}\, I_{1,b} = g^{ab}\, I_{2,b} = g^{ab}\, I_{3,b} = 0 \,, \\
	\label{Universality-Comp-3} 
	& b^{ab}\, I_{1,b} = b^{ab}\, I_{2,b} = b^{ab}\, I_{3,b} = 0 \,, \\
	\label{Universality-Comp-4} 
	& c^{ab}\, I_{1,b} = c^{ab}\, I_{2,b} = c^{ab}\, I_{3,b} = 0 \,.
\end{empheq}
The universality constraints \eqref{Universality-Comp-2} imply that $I_1, I_2, I_3$ are constant. 
Consequently, the conditions \eqref{Universality-Comp-3} and \eqref{Universality-Comp-4} are automatically satisfied. 
In summary,
\begin{equation} \label{Universality-Cauchy-Eigenstrains}
	I_1, I_2, I_3~\text{are constant}, \qquad 
	b^{ab}{}_{|b}=c^{ab}{}_{|b}=0 \,.
\end{equation}
These are exactly the universality conditions obtained by \citet{YavariGoriely2016} for hyperelasticity. 
Recall that the principal invariants explicitly depend on eigenstrains through the material metric $\mathbf{G}$; for instance, $I_1 = \operatorname{tr}\mathbf{b} = b^{ab} g_{ab} = F^a{}_A\,F^b{}_B\,G^{AB} g_{ab}$. 
It was shown in \citep{YavariGoriely2016} that \eqref{Universality-Cauchy-Eigenstrains} imply that universal deformations are covariantly homogeneous, and for a simply-connected body this means the universal eigenstrains are zero-stress (impotent).  Here we observe that the same conclusion holds when, in the absence of eigenstrains, the body is composed of a Cauchy elastic solid.

\section{Universal Deformations and Residual Stresses in Compressible Isotropic Cauchy Elasticity} \label{C-Elasticity-UC}

For a residually-stressed Cauchy elastic body, equilibrium equations in the absence of body forces read $\operatorname{div}\boldsymbol{\sigma}=\mathbf{0}$.
Substituting \eqref{Cauchy-Stress-Representation} into the equilibrium equations, one obtains
\begin{equation} \label{Equilibrium-RSCE}
\begin{aligned}
	\operatorname{div}\boldsymbol{\sigma} &= \nabla \alpha_0
	+ \alpha_1\, \operatorname{div}\mathbf{b}^\sharp
	+ \alpha_2\, \operatorname{div}\mathbf{c}^\sharp
	+ \mathbf{b}^\sharp \cdot \nabla \alpha_1
	+ \mathbf{c}^\sharp \cdot \nabla \alpha_2
	+ \alpha_3\, \operatorname{div}\mathring{\boldsymbol{\sigma}}
	+ \mathring{\boldsymbol{\sigma}} \cdot \nabla \alpha_3 \\
	&\quad
	+ \alpha_4\, \operatorname{div}\mathring{\boldsymbol{\sigma}}^2
	+ \mathring{\boldsymbol{\sigma}}^2 \cdot \nabla \alpha_4 
	+ \alpha_5\, \operatorname{div}\!\left(\mathbf{b}\, \mathring{\boldsymbol{\sigma}} 
	+ \mathring{\boldsymbol{\sigma}}\, \mathbf{b}\right)
	+ \left(\mathbf{b}\, \mathring{\boldsymbol{\sigma}} + \mathring{\boldsymbol{\sigma}}\,\mathbf{b}\right) \cdot \nabla \alpha_5 \\
	&\quad
	+ \alpha_6\, \operatorname{div}\!\left(\mathbf{c}\,\mathring{\boldsymbol{\sigma}}
	+\mathring{\boldsymbol{\sigma}}\, \mathbf{c}\right)
	+ \left(\mathbf{c}\, \mathring{\boldsymbol{\sigma}} + \mathring{\boldsymbol{\sigma}}\, \mathbf{c}\right) \cdot \nabla \alpha_6\\
	&\quad 
	+ \alpha_7\, \operatorname{div}\!\left(\mathbf{b}\, \mathring{\boldsymbol{\sigma}}^2 
	+ \mathring{\boldsymbol{\sigma}}^2\, \mathbf{b}\right)
	+ \left(\mathbf{b}\, \mathring{\boldsymbol{\sigma}}^2 
	+ \mathring{\boldsymbol{\sigma}}^2\, \mathbf{b}\right) \cdot \nabla \alpha_7
	= \mathbf{0}\,.
\end{aligned}
\end{equation}
Notice that, in general, $\operatorname{div}\mathring{\boldsymbol{\sigma}} \neq \mathbf{0}$.

We aim to characterize all deformations that can be maintained in the absence of body forces. That is, we seek pairs of deformations and residual stresses for which the equilibrium equations admit a solution for arbitrary strain-energy density functions. We note that such pairs of deformations and residual stresses solve the equilibrium equations for any isotropic Cauchy elastic solid. This implies that the constitutive response functions---and their derivatives---can be chosen arbitrarily.
In particular, we must have $\nabla \alpha_0=\mathbf{0}$. But observe that
\begin{equation}
	\nabla\alpha_0=\sum_{j=1}^{10} \frac{\partial \alpha_0}{\partial I_j}\,\nabla I_j=\mathbf{0}\,.
\end{equation}
Therefore, the arbitrariness of the partial derivatives of $\alpha_0$ with respect to the nine invariants implies that
\begin{equation} \label{Invarinats-Constant}
	I_1,\hdots,I_{10}~\text{~are~constant}\,.
\end{equation}
Now the equilibrium equations \eqref{Equilibrium-RSCE} are simplified to read
\begin{equation}  \label{Equilibrium-RSCE1}
\begin{aligned}
	& \alpha_1\, \operatorname{div}\mathbf{b}^\sharp
	+ \alpha_2\, \operatorname{div}\mathbf{c}^\sharp
	+ \alpha_3\, \operatorname{div}\mathring{\boldsymbol{\sigma}} 
	+ \alpha_4\, \operatorname{div}\mathring{\boldsymbol{\sigma}}^2
	+ \alpha_5\, \operatorname{div}\!\left(\mathbf{b}\, \mathring{\boldsymbol{\sigma}} 
	+ \mathring{\boldsymbol{\sigma}}\, \mathbf{b}\right) \\
	&\quad
	+ \alpha_6\, \operatorname{div}\!\left(\mathbf{c}\,\mathring{\boldsymbol{\sigma}}
	+\mathring{\boldsymbol{\sigma}}\, \mathbf{c}\right)
	+ \alpha_7\, \operatorname{div}\!\left(\mathbf{b}\, \mathring{\boldsymbol{\sigma}}^2 
	+ \mathring{\boldsymbol{\sigma}}^2\, \mathbf{b}\right)
	= \mathbf{0}\,.
\end{aligned}
\end{equation}
In \eqref{Equilibrium-RSCE}, as $\alpha_1$ and $\alpha_2$ can be chosen arbitrarily, their coefficients must vanish, and hence $\operatorname{div}\mathbf{b}^\sharp=\operatorname{div}\mathbf{c}^\sharp=\mathbf{0}$. These together with knowing that $I_1$, $I_2$, and $I_3$ are constant imply that universal deformations must be homogeneous \citep{Ericksen1955}.
Knowing that $\alpha_3$ can be chosen arbitrarily, its coefficient must vanish, i.e., $\operatorname{div}\mathring{\boldsymbol{\sigma}}=\mathbf{0}$. Constancy of $I_4$, $I_5$, and $I_6$ and knowing that $\operatorname{div}\mathring{\boldsymbol{\sigma}}=\mathbf{0}$ implies that $\mathring{\boldsymbol{\sigma}}$ is homogeneous \citep[Lemma~3.1]{Yavari2024ImplicitElasticity}. 
Knowing that $\mathbf{b}^\sharp$, $\mathbf{c}^\sharp$, and $\mathring{\boldsymbol{\sigma}}$ are homogeneous, equilibrium equations \eqref{Equilibrium-RSCE1} are now trivially satisfied.

Recall that $\mathring{\mathbf{S}}=\mathbf{F}^{-1}\mathring{\boldsymbol{\sigma}}\mathbf{F}^{-\star}$. Knowing that both $\mathbf{F}$ and $\mathring{\boldsymbol{\sigma}}$ are homogeneous, one concludes that the residual stress field $\mathring{\mathbf{S}}$ is homogeneous.
As mentioned earlier, a residual stress field must be divergence-free and satisfy traction-free boundary conditions. These requirements preclude uniformity, implying that non-trivial universal residual stresses cannot exist. Thus, we have proved the following result.

\begin{prop}
For homogeneous compressible isotropic Cauchy elastic solids with residual stress, the set of universal deformations consists precisely of all homogeneous deformations. Moreover, there are no non-trivial universal residual stresses.
\end{prop}

\begin{remark}
\citet{YavariGoriely2016} studied universal deformations in compressible hyperelastic bodies with finite eigenstrains, assuming isotropy in the absence of eigenstrains. In two dimensions, they showed that the material manifold must be flat, implying that universal eigenstrains must be zero-stress. In three dimensions, they proved that the material manifold must be a Riemannian symmetric space and, under the assumption of simple connectivity, flat. Thus, in dimensions two and three, the only universal eigenstrains in simply-connected bodies are zero-stress, and all universal deformations are homogeneous.
In this paper, we do not specify the source of residual stress. However, our result is consistent with that of \citet{YavariGoriely2016}: universal deformations are homogeneous, and no non-trivial universal residual stresses exist.
\end{remark}

\section{Conclusions}  \label{Sec:Conclusions}

In this paper we investigated universal deformations in compressible isotropic Cauchy elastic solids with residual stress, without specifying the source of the residual stress. We showed that universal deformations are homogeneous, and the associated residual stresses must also be homogeneous. Since a non-trivial residual stress cannot be homogeneous, it follows that residual stress must vanish. Thus, a compressible Cauchy elastic solid with a non-trivial distribution of residual stress cannot admit universal deformations. This result is consistent with \citet{YavariGoriely2016}, who proved that in the presence of eigenstrains, universal deformations are covariantly homogeneous and the universal eigenstrains are zero-stress (impotent).

\bibliographystyle{abbrvnat}
\bibliography{ref,ref1}

\begin{thebibliography}{91}
\providecommand{\natexlab}[1]{#1}
\providecommand{\url}[1]{\texttt{#1}}
\expandafter\ifx\csname urlstyle\endcsname\relax
  \providecommand{\doi}[1]{doi: #1}\else
  \providecommand{\doi}{doi: \begingroup \urlstyle{rm}\Url}\fi

\bibitem[Abraham et~al.(2012)Abraham, Marsden, and Ratiu]{Abraham2012Manifolds}
R.~Abraham, J.~E. Marsden, and T.~Ratiu.
\newblock \emph{Manifolds, Tensor Analysis, and Applications}, volume~75.
\newblock Springer Science \& Business Media, 2012.

\bibitem[Beatty(1978)]{Beatty1978}
M.~F. Beatty.
\newblock General solutions in the equilibrium theory of inextensible elastic
  materials.
\newblock \emph{Acta Mechanica}, 29\penalty0 (1):\penalty0 119--126, 1978.

\bibitem[Beatty(1989)]{Beatty1989}
M.~F. Beatty.
\newblock A class of universal relations for constrained, isotropic elastic
  materials.
\newblock \emph{Acta Mechanica}, 80\penalty0 (3):\penalty0 299--312, 1989.

\bibitem[Beskos(1972)]{Beskos1972}
D.~E. Beskos.
\newblock Universal solutions for fiber-reinforced compressible isotropic
  elastic materials.
\newblock \emph{Journal of Elasticity}, 2\penalty0 (3):\penalty0 153--168,
  1972.

\bibitem[Beskos(1973)]{Beskos1973}
D.~E. Beskos.
\newblock Universal solutions for fiber-reinforced incompressible isotropic
  elastic materials.
\newblock \emph{International Journal of Solids and Structures}, 9\penalty0
  (4):\penalty0 553--567, 1973.

\bibitem[Biot(1939)]{Biot1939}
M.~A. Biot.
\newblock Non-linear theory of elasticity and the linearized case for a body
  under initial stress.
\newblock \emph{Philosophical Magazine}, 27\penalty0 (183):\penalty0 468--489,
  1939.

\bibitem[Boehler(1977)]{Boehler1977}
J.-P. Boehler.
\newblock On irreducible representations for isotropic scalar functions.
\newblock \emph{Zeitschrift f{\"u}r Angewandte Mathematik und Mechanik},
  57\penalty0 (6):\penalty0 323--327, 1977.

\bibitem[Boehler(1987)]{Boehler1987}
J.-P. Boehler.
\newblock Representations for isotropic and anisotropic non-polynomial tensor
  functions.
\newblock In \emph{Applications of Tensor Functions in Solid Mechanics}, pages
  31--53. Springer, 1987.

\bibitem[Carroll(1973)]{Carroll1973}
M.~M. Carroll.
\newblock Controllable states of stress for compressible elastic solids.
\newblock \emph{Journal of Elasticity}, 3:\penalty0 57--61, 1973.

\bibitem[Cauchy(1828)]{Cauchy1828}
A.-L. Cauchy.
\newblock Sur les {\'e}quations qui expriment les conditions d'{\'e}quilibre ou
  les lois du mouvement int{\'e}rieur d'un corps solide, {\'e}lastique ou non
  {\'e}lastique.
\newblock \emph{Exercises de Math{\'e}matiques}, 3:\penalty0 160--187, 1828.

\bibitem[Chi et~al.(2015)Chi, Talischi, Lopez-Pamies, and H.~Paulino]{Chi2015}
H.~Chi, C.~Talischi, O.~Lopez-Pamies, and G.~H.~Paulino.
\newblock Polygonal finite elements for finite elasticity.
\newblock \emph{International Journal for Numerical Methods in Engineering},
  101\penalty0 (4):\penalty0 305--328, 2015.

\bibitem[De~Tommasi(1996)]{Tommasi1996}
D.~De~Tommasi.
\newblock Elastic bodies reinforced with inextensible surfaces.
\newblock \emph{Journal of Elasticity}, 45\penalty0 (3):\penalty0 215--250,
  1996.

\bibitem[Dragoni(1996)]{Dragoni1996}
E.~Dragoni.
\newblock The radial compaction of a hyperelastic tube as a benchmark in
  compressible finite elasticity.
\newblock \emph{International Journal of Non-Linear Mechanics}, 31\penalty0
  (4):\penalty0 483--493, 1996.

\bibitem[Eckart(1948)]{Eckart1948}
C.~Eckart.
\newblock The thermodynamics of irreversible processes. 4. {T}he theory of
  elasticity and anelasticity.
\newblock \emph{Physical Review}, 73\penalty0 (4):\penalty0 373--382, 1948.

\bibitem[Ericksen(1954)]{Ericksen1954}
J.~L. Ericksen.
\newblock Deformations possible in every isotropic, incompressible, perfectly
  elastic body.
\newblock \emph{Zeitschrift f{\"u}r Angewandte Mathematik und Physik},
  5\penalty0 (6):\penalty0 466--489, 1954.

\bibitem[Ericksen(1955)]{Ericksen1955}
J.~L. Ericksen.
\newblock Deformations possible in every compressible, isotropic, perfectly
  elastic material.
\newblock \emph{Studies in Applied Mathematics}, 34\penalty0 (1-4):\penalty0
  126--128, 1955.

\bibitem[Fosdick(1966)]{Fosdick1966}
R.~L. Fosdick.
\newblock Remarks on {C}ompatibility.
\newblock \emph{Modern Developments in the Mechanics of Continua}, pages
  109--127, 1966.

\bibitem[Fosdick(1971)]{Fosdick1971}
R.~L. Fosdick.
\newblock Statically possible radially symmetric deformations in isotropic,
  incompressible elastic solids.
\newblock \emph{Zeitschrift f{\"u}r angewandte Mathematik und Physik},
  22:\penalty0 590--607, 1971.

\bibitem[Fosdick and Schuler(1969)]{Fosdick1969}
R.~L. Fosdick and K.~W. Schuler.
\newblock On {E}ricksen's problem for plane deformations with uniform
  transverse stretch.
\newblock \emph{International Journal of Engineering Science}, 7\penalty0
  (2):\penalty0 217--233, 1969.

\bibitem[Gairola(1979)]{Gairola1979}
B.~Gairola.
\newblock \emph{Nonlinear elastic problems}.
\newblock F.R.N. Nabarro (Ed.), Dislocations in Solids. North-Holland
  Publishing Co., Amsterdam, 1979.

\bibitem[Golgoon and Yavari(2018{\natexlab{a}})]{Golgoon2017}
A.~Golgoon and A.~Yavari.
\newblock Nonlinear elastic inclusions in anisotropic solids.
\newblock \emph{Journal of Elasticity}, 130\penalty0 (2):\penalty0 239--269,
  2018{\natexlab{a}}.

\bibitem[Golgoon and Yavari(2018{\natexlab{b}})]{Golgoon2018}
A.~Golgoon and A.~Yavari.
\newblock Line and point defects in nonlinear anisotropic solids.
\newblock \emph{Zeitschrift f{\"u}r angewandte Mathematik und Physik},
  69:\penalty0 1--28, 2018{\natexlab{b}}.

\bibitem[Golgoon and Yavari(2021)]{Golgoon2021}
A.~Golgoon and A.~Yavari.
\newblock On {H}ashin’s hollow cylinder and sphere assemblages in anisotropic
  nonlinear elasticity.
\newblock \emph{Journal of Elasticity}, 146\penalty0 (1):\penalty0 65--82,
  2021.

\bibitem[Goodbrake et~al.(2020)Goodbrake, Yavari, and Goriely]{Goodbrake2020}
C.~Goodbrake, A.~Yavari, and A.~Goriely.
\newblock The anelastic {E}ricksen problem: {U}niversal deformations and
  universal eigenstrains in incompressible nonlinear anelasticity.
\newblock \emph{Journal of Elasticity}, 142\penalty0 (2):\penalty0 291--381,
  2020.

\bibitem[Goriely(2017)]{Goriely2017}
A.~Goriely.
\newblock \emph{The Mathematics and Mechanics of Biological Growth}, volume~45.
\newblock Springer, 2017.

\bibitem[Gurtin(1972)]{Gurtin1972}
M.~E. Gurtin.
\newblock The linear theory of elasticity.
\newblock In \emph{Handbuch der Physik, Band VIa/2.} Springer-Verlag, Berlin,
  1972.

\bibitem[Hashin(1985)]{Hashin1985}
Z.~Hashin.
\newblock Large isotropic elastic deformation of composites and porous media.
\newblock \emph{International Journal of Solids and Structures}, 21\penalty0
  (7):\penalty0 711--720, 1985.

\bibitem[Hoger(1985)]{Hoger1985}
A.~Hoger.
\newblock On the residual stress possible in an elastic body with material
  symmetry.
\newblock \emph{Archive for Rational Mechanics and Analysis}, 88\penalty0
  (3):\penalty0 271--289, 1985.

\bibitem[Johnson and Hoger(1995)]{JohnsonHoger1995}
B.~E. Johnson and A.~Hoger.
\newblock The use of a virtual configuration in formulating constitutive
  equations for residually stressed elastic materials.
\newblock \emph{Journal of Elasticity}, 41\penalty0 (3):\penalty0 177--215,
  1995.

\bibitem[Kafadar(1972)]{Kafadar1972}
C.~B. Kafadar.
\newblock On {E}ricksen's problem.
\newblock \emph{Archive for Rational Mechanics and Analysis}, 47:\penalty0
  15--27, 1972.

\bibitem[Klingbeil and Shield(1966)]{KlingbeilShield1966}
W.~W. Klingbeil and R.~T. Shield.
\newblock On a class of solutions in plane finite elasticity.
\newblock \emph{Zeitschrift f{\"u}r angewandte Mathematik und Physik},
  17\penalty0 (4):\penalty0 489--511, 1966.

\bibitem[Knowles(1979)]{Knowles1979}
J.~K. Knowles.
\newblock Universal states of finite anti-plane shear: {E}ricksen's problem in
  miniature.
\newblock \emph{The American Mathematical Monthly}, 86\penalty0 (2):\penalty0
  109--113, 1979.

\bibitem[Kondo(1949)]{Kondo1949}
K.~Kondo.
\newblock A proposal of a new theory concerning the yielding of materials based
  on {R}iemannian geometry.
\newblock \emph{The Journal of the Japan Society of Aeronautical Engineering},
  2\penalty0 (8):\penalty0 29--31, 1949.

\bibitem[Kurashige(1985)]{Kurashige1985}
M.~Kurashige.
\newblock Finite deformations of an area-preserving material.
\newblock \emph{Zeitschrift f{\"u}r angewandte Mathematik und Physik},
  36\penalty0 (6):\penalty0 822--836, 1985.

\bibitem[Lee and Bhattacharya(2023)]{LeeBhattacharya2023}
V.~Lee and K.~Bhattacharya.
\newblock Universal deformations of incompressible nonlinear elasticity as
  applied to ideal liquid crystal elastomers.
\newblock \emph{Journal of Elasticity}, pages 1--27, 2023.

\bibitem[Lopez-Pamies et~al.(2012)Lopez-Pamies, Moraleda, Segurado, and
  Llorca]{Lopez2012}
O.~Lopez-Pamies, J.~Moraleda, J.~Segurado, and J.~Llorca.
\newblock On the extremal properties of {H}ashin’s hollow cylinder assemblage
  in nonlinear elasticity.
\newblock \emph{Journal of Elasticity}, 107:\penalty0 1--10, 2012.

\bibitem[Marris(1975)]{Marris1975}
A.~Marris.
\newblock Universal deformations in incompressible isotropic elastic materials.
\newblock \emph{Journal of Elasticity}, 5\penalty0 (2):\penalty0 111--128,
  1975.

\bibitem[Marris(1982)]{Marris1982}
A.~Marris.
\newblock Two new theorems on {E}ricksen's problem.
\newblock \emph{Archive for Rational Mechanics and Analysis}, 79:\penalty0
  131--173, 1982.

\bibitem[Marris and Shiau(1970)]{Marris1970}
A.~Marris and J.~Shiau.
\newblock Universal deformations in isotropic incompressible hyperelastic
  materials when the deformation tensor has equal proper values.
\newblock \emph{Archive for Rational Mechanics and Analysis}, 36\penalty0
  (2):\penalty0 135--160, 1970.

\bibitem[Marsden and Hughes(1994)]{MarsdenHughes1994}
J.~E. Marsden and T.~J.~R. Hughes.
\newblock \emph{Mathematical Foundations of Elasticity}.
\newblock Dover Publications, New York, 1994.

\bibitem[Merodio and Ogden(2016)]{MerodioOgden2016}
J.~Merodio and R.~W. Ogden.
\newblock Extension, inflation and torsion of a residually stressed circular
  cylindrical tube.
\newblock \emph{Continuum Mechanics and Thermodynamics}, 28\penalty0
  (1):\penalty0 157--174, 2016.

\bibitem[Merodio et~al.(2013)Merodio, Ogden, and
  Rodríguez]{MerodioOgdenRodriguez2013}
J.~Merodio, R.~W. Ogden, and J.~Rodríguez.
\newblock The influence of residual stress on finite deformation elastic
  response.
\newblock \emph{International Journal of Non-Linear Mechanics}, 56:\penalty0
  43--49, 2013.

\bibitem[Mihai and Goriely(2023)]{MihaiGoriely2023}
L.~A. Mihai and A.~Goriely.
\newblock Controllable deformations of unconstrained ideal nematic elastomers.
\newblock \emph{Journal of Elasticity}, pages 1--12, 2023.

\bibitem[Morgan(1966)]{Morgan1966}
A.~J.~A. Morgan.
\newblock Some properties of media defined by constitutive equations in
  implicit form.
\newblock \emph{International Journal of Engineering Science}, 4\penalty0
  (2):\penalty0 155--178, 1966.

\bibitem[Ogden(1984)]{Ogden1984}
R.~W. Ogden.
\newblock \emph{Non-Linear Elastic Deformations}.
\newblock Dover, 1984.

\bibitem[Pradhan and Yavari(2023)]{PradhanYavari2023}
S.~P. Pradhan and A.~Yavari.
\newblock Accretion--ablation mechanics.
\newblock \emph{Philosophical Transactions of the Royal Society A},
  381\penalty0 (2263):\penalty0 20220373, 2023.

\bibitem[Rajagopal(2003)]{Rajagopal2003}
K.~R. Rajagopal.
\newblock On implicit constitutive theories.
\newblock \emph{Applications of Mathematics}, 48:\penalty0 279--319, 2003.

\bibitem[Rajagopal(2007)]{Rajagopal2007}
K.~R. Rajagopal.
\newblock The elasticity of elasticity.
\newblock \emph{Zeitschrift f{\"u}r angewandte Mathematik und Physik},
  58:\penalty0 309--317, 2007.

\bibitem[Rivlin(1948)]{Rivlin1948}
R.~S. Rivlin.
\newblock Large elastic deformations of isotropic materials {IV}. {F}urther
  developments of the general theory.
\newblock \emph{Philosophical Transactions of the Royal Society of London A},
  241\penalty0 (835):\penalty0 379--397, 1948.

\bibitem[Rivlin(1949{\natexlab{a}})]{Rivlin1949a}
R.~S. Rivlin.
\newblock Large elastic deformations of isotropic materials. {V}. {T}he problem
  of flexure.
\newblock \emph{Proceedings of the Royal Society of London A}, 195\penalty0
  (1043):\penalty0 463--473, 1949{\natexlab{a}}.

\bibitem[Rivlin(1949{\natexlab{b}})]{Rivlin1949b}
R.~S. Rivlin.
\newblock A note on the torsion of an incompressible highly elastic cylinder.
\newblock In \emph{Mathematical Proceedings of the Cambridge Philosophical
  Society}, volume~45, pages 485--487. Cambridge University Press,
  1949{\natexlab{b}}.

\bibitem[Rivlin and Ericksen(1955)]{RivlinEricksen1955}
R.~S. Rivlin and J.~L. Ericksen.
\newblock Stress-deformation relations for isotropic materials.
\newblock \emph{Journal of Rational Mechanics and Analysis}, 4:\penalty0
  323--425, 1955.

\bibitem[Rivlin and Saunders(1951)]{Rivlin1951}
R.~S. Rivlin and D.~W. Saunders.
\newblock Large elastic deformations of isotropic materials {VII}.
  {E}xperiments on the deformation of rubber.
\newblock \emph{Philosophical Transactions of the Royal Society of London A},
  243\penalty0 (865):\penalty0 251--288, 1951.

\bibitem[Saccomandi(2001)]{Saccomandi2001}
G.~Saccomandi.
\newblock Universal solutions and relations in finite elasticity.
\newblock In \emph{Topics in finite elasticity}, pages 95--130. Springer, 2001.

\bibitem[Sadik and Yavari(2017)]{SadikYavari2017}
S.~Sadik and A.~Yavari.
\newblock On the origins of the idea of the multiplicative decomposition of the
  deformation gradient.
\newblock \emph{Mathematics and Mechanics of Solids}, 22:\penalty0 771--772,
  2017.

\bibitem[Shams et~al.(2011)Shams, Destrade, and Ogden]{Shams2011}
M.~Shams, M.~Destrade, and R.~W. Ogden.
\newblock Initial stresses in elastic solids: {C}onstitutive laws and
  acoustoelasticity.
\newblock \emph{Wave Motion}, 48\penalty0 (7):\penalty0 552--567, 2011.

\bibitem[Shojaei and Yavari(2018)]{Shojaei2018}
M.~F. Shojaei and A.~Yavari.
\newblock Compatible-strain mixed finite element methods for incompressible
  nonlinear elasticity.
\newblock \emph{Journal of Computational Physics}, 361:\penalty0 247--279,
  2018.

\bibitem[Singh and Pipkin(1965)]{SinghPipkin1965}
M.~Singh and A.~C. Pipkin.
\newblock Note on {E}ricksen's problem.
\newblock \emph{Zeitschrift f{\"u}r angewandte Mathematik und Physik},
  16\penalty0 (5):\penalty0 706--709, 1965.

\bibitem[Spencer(1970)]{Spencer1970}
A.~J.~M. Spencer.
\newblock A note on the decomposition of tensors into traceless symmetric
  tensors.
\newblock \emph{International Journal of Engineering Science}, 8\penalty0
  (6):\penalty0 475--481, 1970.
\newblock \doi{10.1016/0020-7225(70)90031-5}.

\bibitem[Spencer(1971)]{Spencer1971}
A.~J.~M. Spencer.
\newblock Part {III}. {T}heory of invariants.
\newblock \emph{Continuum Physics}, 1:\penalty0 239--353, 1971.

\bibitem[Tadmor et~al.(2012)Tadmor, Miller, and Elliott]{Tadmor2012}
E.~B. Tadmor, R.~E. Miller, and R.~S. Elliott.
\newblock \emph{Continuum Mechanics and Thermodynamics: From Fundamental
  Concepts to Governing Equations}.
\newblock Cambridge University Press, 2012.

\bibitem[Truesdell(1952)]{Truesdell1952}
C.~Truesdell.
\newblock The mechanical foundations of elasticity and fluid dynamics.
\newblock \emph{Journal of Rational Mechanics and Analysis}, 1\penalty0
  (1):\penalty0 125--300, 1952.

\bibitem[Truesdell(1966)]{Truesdell1966}
C.~Truesdell.
\newblock \emph{The Elements of Continuum Mechanics}.
\newblock Springer-Verlag, 1966.

\bibitem[Truesdell and Noll(2004)]{TruesdellNoll2004}
C.~Truesdell and W.~Noll.
\newblock \emph{The Non-Linear Field Theories of Mechanics}.
\newblock Springer, 2004.

\bibitem[Wang(1969)]{Wang1969}
C.~C. Wang.
\newblock On representations for isotropic functions: {P}art {I}. {I}sotropic
  functions of symmetric tensors and vectors.
\newblock \emph{Archive for Rational Mechanics and Analysis}, 33:\penalty0
  249--267, 1969.

\bibitem[Wesolowski and Seeger(1968)]{Wesolowski1968}
Z.~Wesolowski and A.~Seeger.
\newblock On the screw dislocation in finite elasticity.
\newblock In \emph{Mechanics of Generalized Continua. Proceedings of the IUTAM
  Symposium on the Generalized Cosserat Continuum and the Continuum Theory of
  Dislocations with Applications (Ed. E. Kr{\"o}ner, Springer, Berlin, etc.,
  1968)}, pages 294--297. Springer, 1968.

\bibitem[Yavari(2021{\natexlab{a}})]{Yavari2021}
A.~Yavari.
\newblock Universal deformations in inhomogeneous isotropic nonlinear elastic
  solids.
\newblock \emph{Proceedings of the Royal Society A}, 477\penalty0
  (2253):\penalty0 20210547, 2021{\natexlab{a}}.

\bibitem[Yavari(2021{\natexlab{b}})]{Yavari2021Eshelby}
A.~Yavari.
\newblock On {E}shelby’s inclusion problem in nonlinear anisotropic
  elasticity.
\newblock \emph{Journal of Micromechanics and Molecular Physics}, 6\penalty0
  (01):\penalty0 2150002, 2021{\natexlab{b}}.

\bibitem[Yavari(2023)]{Yavari2023}
A.~Yavari.
\newblock Universal displacements in inextensible fiber-reinforced linear
  elastic solids.
\newblock \emph{Mathematics and Mechanics of Solids}, pages 1--17, 2023.
\newblock \doi{10.1177/10812865231181924}.

\bibitem[Yavari(2024)]{Yavari2024Cauchy}
A.~Yavari.
\newblock Universal deformations and inhomogeneities in isotropic {C}auchy
  elasticity.
\newblock \emph{Proceedings of the Royal Society A}, 480\penalty0
  (2277):\penalty0 20240229, 2024.

\bibitem[Yavari(2025)]{Yavari2025Fibers}
A.~Yavari.
\newblock On universal deformations of compressible {C}auchy elastic solids
  reinforced by inextensible fibers.
\newblock \emph{arXiv preprint arXiv:2506.11203}, June 2025.
\newblock URL \url{https://arxiv.org/abs/2506.11203}.

\bibitem[Yavari and Golgoon(2019)]{Yavari2019Cloaking}
A.~Yavari and A.~Golgoon.
\newblock Nonlinear and linear elastodynamic transformation cloaking.
\newblock \emph{Archive for Rational Mechanics and Analysis}, 234\penalty0
  (1):\penalty0 211--316, 2019.

\bibitem[Yavari and Goriely(2012{\natexlab{a}})]{YavariGoriely2012a}
A.~Yavari and A.~Goriely.
\newblock Riemann--{C}artan geometry of nonlinear dislocation mechanics.
\newblock \emph{Archive for Rational Mechanics and Analysis}, 205\penalty0
  (1):\penalty0 59--118, 2012{\natexlab{a}}.

\bibitem[Yavari and Goriely(2012{\natexlab{b}})]{YavariGoriely2012b}
A.~Yavari and A.~Goriely.
\newblock Weyl geometry and the nonlinear mechanics of distributed point
  defects.
\newblock \emph{Proceedings of the Royal Society A}, 468\penalty0
  (2148):\penalty0 3902--3922, 2012{\natexlab{b}}.

\bibitem[Yavari and Goriely(2013{\natexlab{a}})]{YavariGoriely2013a}
A.~Yavari and A.~Goriely.
\newblock Riemann--{C}artan geometry of nonlinear disclination mechanics.
\newblock \emph{Mathematics and Mechanics of Solids}, 18\penalty0 (1):\penalty0
  91--102, 2013{\natexlab{a}}.

\bibitem[Yavari and Goriely(2013{\natexlab{b}})]{YavariGoriely2013b}
A.~Yavari and A.~Goriely.
\newblock Nonlinear elastic inclusions in isotropic solids.
\newblock \emph{Proceedings of the Royal Society A}, 469\penalty0
  (2160):\penalty0 20130415, 2013{\natexlab{b}}.

\bibitem[Yavari and Goriely(2014)]{YavariGoriely2014}
A.~Yavari and A.~Goriely.
\newblock The geometry of discombinations and its applications to semi-inverse
  problems in anelasticity.
\newblock \emph{Proceedings of the Royal Society A}, 470\penalty0
  (2169):\penalty0 20140403, 2014.

\bibitem[Yavari and Goriely(2015)]{YavariGoriely2015}
A.~Yavari and A.~Goriely.
\newblock The twist-fit problem: {F}inite torsional and shear eigenstrains in
  nonlinear elastic solids.
\newblock \emph{Proceedings of the Royal Society A}, 471\penalty0
  (2183):\penalty0 20150596, 2015.

\bibitem[Yavari and Goriely(2016)]{YavariGoriely2016}
A.~Yavari and A.~Goriely.
\newblock The anelastic {E}ricksen problem: {U}niversal eigenstrains and
  deformations in compressible isotropic elastic solids.
\newblock \emph{Proceedings of the Royal Society A}, 472\penalty0
  (2196):\penalty0 20160690, 2016.

\bibitem[Yavari and Goriely(2021)]{YavariGoriely2021}
A.~Yavari and A.~Goriely.
\newblock Universal deformations in anisotropic nonlinear elastic solids.
\newblock \emph{Journal of the Mechanics and Physics of Solids}, 156:\penalty0
  104598, 2021.

\bibitem[Yavari and
  Goriely(2022{\natexlab{a}})]{Yavari2022Anelastic-Universality}
A.~Yavari and A.~Goriely.
\newblock Universality in anisotropic linear anelasticity.
\newblock \emph{Journal of Elasticity}, 150\penalty0 (2):\penalty0 241--259,
  2022{\natexlab{a}}.

\bibitem[Yavari and Goriely(2022{\natexlab{b}})]{YavariGoriely2022}
A.~Yavari and A.~Goriely.
\newblock The universal program of linear elasticity.
\newblock \emph{Mathematics and Mechanics of Solids}, 2022{\natexlab{b}}.

\bibitem[Yavari and Goriely(2023)]{YavariGoriely2023Universal}
A.~Yavari and A.~Goriely.
\newblock The universal program of nonlinear hyperelasticity.
\newblock \emph{Journal of Elasticity}, 154\penalty0 (1):\penalty0 91--146,
  2023.

\bibitem[Yavari and Goriely(2024)]{Yavari2024ImplicitElasticity}
A.~Yavari and A.~Goriely.
\newblock Controllable deformations in compressible isotropic implicit
  elasticity.
\newblock \emph{Zeitschrift f{\"u}r angewandte Mathematik und Physik},
  75\penalty0 (5):\penalty0 169, 2024.

\bibitem[Yavari and Goriely(2025)]{YavariGoriely2025}
A.~Yavari and A.~Goriely.
\newblock Nonlinear {C}auchy elasticity.
\newblock \emph{Archive for Rational Mechanics and Analysis}, 2025.
\newblock \doi{10.1007/s00205-025-02120-0}.

\bibitem[Yavari and Pradhan(2022)]{YavariPradhan2022}
A.~Yavari and S.~P. Pradhan.
\newblock Accretion mechanics of nonlinear elastic circular cylindrical bars
  under finite torsion.
\newblock \emph{Journal of Elasticity}, 152\penalty0 (1-2):\penalty0 29--60,
  2022.

\bibitem[Yavari and Sfyris(2025)]{YavariSfyris2025}
A.~Yavari and D.~Sfyris.
\newblock Universal displacements in anisotropic linear {C}auchy elasticity.
\newblock \emph{Journal of Elasticity}, 157\penalty0 (1):\penalty0 1--15, 2025.

\bibitem[Yavari and Sozio(2023)]{yavariSozio2023}
A.~Yavari and F.~Sozio.
\newblock On the direct and reverse multiplicative decompositions of
  deformation gradient in nonlinear anisotropic anelasticity.
\newblock \emph{Journal of the Mechanics and Physics of Solids}, 170:\penalty0
  105101, 2023.

\bibitem[Yavari et~al.(2020)Yavari, Goodbrake, and Goriely]{Yavari2020}
A.~Yavari, C.~Goodbrake, and A.~Goriely.
\newblock Universal displacements in linear elasticity.
\newblock \emph{Journal of the Mechanics and Physics of Solids}, 135:\penalty0
  103782, 2020.

\bibitem[Yavari et~al.(2023)Yavari, Safa, and
  Soleiman~Fallah]{YavariAccretion2023}
A.~Yavari, Y.~Safa, and A.~Soleiman~Fallah.
\newblock Finite extension of accreting nonlinear elastic solid circular
  cylinders.
\newblock \emph{Continuum Mechanics and Thermodynamics}, pages 1--17, 2023.

\bibitem[Zubov(1997)]{Zubov1997}
L.~M. Zubov.
\newblock \emph{Nonlinear Theory of Dislocations and Disclinations in Elastic
  Bodies}, volume~47.
\newblock Springer Science \& Business Media, 1997.

\end{thebibliography}

\end{document}